\documentclass[twocolumn]{aastex6}
\usepackage{epsfig}
\usepackage{graphicx}
\usepackage{float}
\usepackage{amsmath}
\usepackage{color}
\usepackage{amssymb}
\usepackage{amsfonts}
\usepackage{units}
\begin{document}

\title{Persistent Radio Emission from Synchrotron Heating by a Repeating Fast Radio Burst Source in a Nebula}

\author{Qiao-Chu Li\altaffilmark{1,2}, Yuan-Pei Yang\altaffilmark{3}, and Zi-Gao Dai\altaffilmark{1,2}}

\affil{
$^1$School of Astronomy and Space Science, Nanjing University, Nanjing 210093, China; dzg@nju.edu.cn\\
$^2$ Key laboratory of Modern Astronomy and Astrophysics (Nanjing University), Ministry of Education, Nanjing 210093, China \\
$^3$ South-Western Institute for Astronomy Research, Yunnan University, Kunming, Yunnan, China
}

\begin{abstract}

The first repeating fast radio burst (FRB), FRB 121102, was found to be associated with a spatially coincident, persistent nonthermal radio source, but the origin of the persistent emission remains unknown. In this paper, we propose that the persistent emission is produced via synchrotron-heating process by multiple bursts of FRB 121102 in a self-absorbed synchrotron nebula. As a population of bursts of the repeating FRB absorbed by the synchrotron nebula, the energy distribution of electrons in the nebula will change significantly. As a result, the spectrum of the nebula will show a hump steadily. For the persistent emission of FRB 121102, the total energy of bursts injecting into the nebula is required to be about $3.3\times10^{49}\,\unit{erg}$, the burst injection age is over $6.7\times 10^4\,\unit{yr}$, the nebula size is $\sim0.02\,\unit{pc}$, and the electron number is about $3.2\times10^{55}$. We predict that as more bursts inject, the brightness of the nebula would be brighter than the current observation, and meanwhile, the peak frequency would become higher. Due to the synchrotron absorption of the nebula, some low-frequency bursts would be absorbed, which may explain why most bursts were detected above $\sim1~\unit{GHz}$.
\end{abstract}

\section{Introduction}

Fast radio bursts (FRBs) are mysterious millisecond-duration transients at radio frequency bands ($400\,\unit{MHz}-8\,\unit{GHz}$).
Up to now, above 100 FRBs have been discovered \citep{lor07,kea12,tho13,spi14,mas15,rav15,rav16,cha16,spi16,pet17,cha17,
sha18,chi18,zha19i,ban19}, see \cite{pet16} for a catalog of
published FRBs~\footnote{http://frbcat.org/}.
20 of them show repeating behaviors, including FRB 121102 \citep{spi14,spi16,sch16,cha17}, FRB 180814.J0422+73 \citep{chi19}, FRB 171019 \citep{kum19}, and 17 FRBs recently discovered by the Canadian Hydrogen Intensity Mapping Experiment \citep[CHIME;][]{chi19b,fon20}.
Some observations show that FRBs are of cosmological origin: First, the DM of an FRB is much larger than that contributed by the Milky Way \citep{spi14,pet19}. This implies that the FRBs are extragalactic origin. Second, the host galaxies of five FRBs, including FRB 121102, FRB 180916, FRB 180924, FRB 181112, and FRB 190523 \citep{cha17,mar17,ten17,mar20,ban19,pro19,rav19}, have been directly located. Third, the sky distribution of FRBs is isotropic \citep{opp16,sha18,bha18,jam19,loc19}. Fourth, \cite{sha18} reported the dispersion-brightness relation for FRBs from a wide-field survey using the Australian Square Kilometer Array Pathfinder (ASKAP) with excess DMs characterizing cosmological distances.

The first repeating case is FRB 121102, which was discovered at Arecibo telescope \citep{sch16,spi16}.
\cite{cha17} found that FRB 121102 was coincident with a $0.2\,\unit{mJy}$ persistent radio source achieved from Karl G. Jansky Very Large Array (VLA) observations.
Based on optical imaging and spectroscopy with the Gemini and Keck telescopes, the host galaxy of FRB 121102 was found to be a low-metallicity, star-forming, dwarf galaxy at the redshift $z = 0.193$ \citep{ten17}.
European VLBI Network (EVN) observations further showed that the size of this steady radio synchrotron  source is $\lesssim 0.7 \,\unit{pc}$ and the luminosity is $\nu L_\nu \sim 10^{39}\,\unit{erg\,s^{-1}}$ \citep{mar17}. \cite{mic18} reported almost 100 per cent linearly polarized emission of FRB 121102 with a very high and variable Faraday rotation measure of $\sim10^{5}\,\unit{rad\,m^{-2}}$ corresponding to a characteristic magnetic field strength of about $1\,\unit{mG}$.

So far, only the first repeating FRB, FRB 121102, was found to be associated with a spatially coincident, persistent nonthermal radio source \citep{cha17}.
Some evidence suggested that the luminosity of the persistent radio source of FRB 121102 might be associated with the large rotation measure (RM) \citep{yan20}.
Recently, \cite{eft19} presented the first detection of a radio emission coincident with the super luminous supernova (SLSN) PTF10hgi about 7.5yr post-explosion. The luminosity and the frequency of this radio emission are approximately consistent with the persistent emission of FRB 121102, which implies that there may be some connections between FRBs and SLSNe \citep{met17}.
However, \cite{law19} used the VLA to observe ten type-\uppercase\expandafter{\romannumeral 1} SLSNe at $3\,\unit{GHz}$, and \cite{men19} performed dedicated observations of the remnants of six GRBs using the Arecibo telescope and the Robert C. Byrd Green Bank Telescope (GBT), all without any FRB detected.

The physical origin of repeating FRBs remains unknown. Suggested models for repeating FRBs include giant pulses \citep{con16,cor16,lyu16}, magnetic energy release in a pulsar magnetosphere \citep{pop10,kul14,kat16b,met17,kas17},
maser emission in an outflow \citep{mur16,bel17,met19}, an asteroid belt interacting with a neutron star \citep{dai16}, cosmic comb \citep{zha17,zha18b,iok20}, the fluctuation in the magnetosphere of a neutron star \citep{yan18,wan20}, and so on.

In some proposed scenarios, FRBs are expected to be located in surrounding nebulae, e.g., pulsar wind nebulae (PWNe), SLSNe and/or long gamma-ray bursts (LGRBs) \citep[e.g.,][]{yang16,mur16,met17,bel17,bel19,dai17,mar18,met19,yangy19,mar19}, and the persistent radio emission source associated with FRB 121102 has been suggested to be explained by such a nebula.
Many models have been proposed to explain the connection between FRB and persistent emission. \cite{yang16} studied a synchrotron-heating process by an FRB embedded in a synchrotron nebula with the frequency of the FRB below the synchrotron self-absorption (SSA) frequency of the nebula. They found that the spectra of the nebula would show a hump after the FRB injection. \cite{mur16} and \cite{oma18} studied the quasi-steady emission from the pulsar wind nebulae (PWNe) associated with pulsar-driven SN/SLSN remnants for a long time after the explosion, which might explain the persistent radio emission of FRB 121102. \cite{met17} presented a scenario that FRB 121102 associated with the birth of a young magnetar is embedded within a young hydrogen-poor SN remnant, and the radio emission can arise from the forward shock interaction between the fastest parts of the SN ejecta and the surrounding stellar progenitor wind. Moreover, \cite{bel17,bel19} considered that the persistent radio source may be heated by magnetic dissipation and internal waves released by the magnetar ejecta. In addition, \cite{dai17} and \cite{yangy19} suggested that the persistent radio emission could be synchrotron emission of a non-relativistic PWN arising from an ultra-relativistic wind of a rapidly rotating strongly magnetized pulsar sweeping up its ambient dense interstellar medium without any surrounding SN ejecta.
Furthermore, \cite{mar18} presented that relativistic thermal electrons heated at the termination shock of the magnetar wind can power the persistent source, where a nebula is surrounding a young flaring magnetar.
\cite{wan19} analyzed the multi-wavelength emission from the nebula powered by an FRB central engine, and they found that with a duty cycle consistent with that observed for FRB 121102 , a sporadically active central engine is required.

In this paper, following \citet{yang16}, we propose that the observed persistent radio emission of FRB 121102 could be generated via synchrotron heating by multiple bursts of FRB 121102 in a surrounding nebula. As radio bursts inject, the electron distribution in the nebula would change gradually due to synchrotron heating process, and the nebula spectrum will finally show a significant hump. Meanwhile, the process of the ``synchrotron external absorption'' would cause low-frequency bursts to be absorbed by the nebula. This paper is organized as follows: In Section \ref{sec2}, we analyze the theory of radio bursts injecting into a nebula. The Numerical calculation of the spectra is presented in Section \ref{sec3}. Finally, the results are summarized with some discussions in Section \ref{sec4}.

\section{Synchrotron heating/external absorption}\label{sec2}

We first summary the physics of synchrotron heating/external absorption within the context of FRB-nebula interaction \citep{yang16}.
We assume that the initial electron number density in the nebula satisfies a power-law distribution before FRB injection, e.g.,
\begin{eqnarray}
N_{\rm e}(\gamma,0)=K\gamma^{-p}, \quad  (\gamma_{\min} \leqslant \gamma \leqslant \gamma_{\max})\label{electrondistribution}
\end{eqnarray}
where $\gamma_{\rm min}$ and $\gamma_{\rm max}$ are the minimum and maximum electron Lorentz factor, respectively. Electrons in the nebula are assumed to have an isotropic distribution of pitch angles relative to the magnetic field $B$. The intensity of SSA in the nebula can be expressed as \citep{ghi13}:
\begin{eqnarray}
I_\nu
&=&\frac{2m_e}{\sqrt{3}\sqrt{\nu_B}}\nu^{5/2}(1-e^{-\tau_\nu})f_I(p),
\end{eqnarray}
with
\begin{align}
f_I(p)&=\dfrac {f_\epsilon(p)}{f_\kappa(p)}=\frac{1}{p+1}\notag\\
&\times \dfrac{\Gamma\left(\dfrac{3p-1}{12}\right)\Gamma\left(\dfrac{3p+19}{12}\right)\Gamma\left(\dfrac{p+5}{4}\right)\Gamma\left(\dfrac{p+8}{4}\right)}{\Gamma\left(\dfrac{3p+2}{12}\right)\Gamma\left(\dfrac{3p+22}{12}\right)\Gamma\left(\dfrac{p+6}{4}\right)\Gamma\left(\dfrac{p+7}{4}\right)},
\end{align}
where $\Gamma(x)$ is the Gamma function, and all the products of the Gamma function are contained in $f_\epsilon(p)$ and $f_\kappa(p)$:
\begin{equation}
f_\epsilon(p)=\dfrac{\Gamma\left(\dfrac{3p-1}{12}\right)\Gamma\left(\dfrac{3p+19}{12}\right)\Gamma\left(\dfrac{p+5}{4}\right)}{(p+1)\Gamma\left(\dfrac{p+7}{4}\right)},
\end{equation}
and
\begin{equation}
f_\kappa(p)=\dfrac{\Gamma\left(\dfrac{3p+2}{12}\right)\Gamma\left(\dfrac{3p+22}{12}\right)\Gamma\left(\dfrac{p+6}{4}\right)}{\Gamma\left(\dfrac{p+8}{4}\right)}.
\end{equation}
The Larmor frequency $\nu_B$ is defined as $\nu_B\equiv eB/(2\pi m_e c)$. The SSA optical depth is
\begin{equation}
\tau_\nu =3^{\frac{p+1}{2}}\frac{\sqrt \pi e^2KR}{8 m_ec}\frac{1}{\nu_B}\left(\frac{\nu}{\nu_B}\right)^{-\frac{p+4}{2}}f_\kappa(p),
\end{equation}
where $R$ is regarded as the thickness of the nebula. The initial SSA frequency $\nu_{a,\,0}$, defined by $\tau_\nu=1$, is
\begin{eqnarray}
\nu_{a,\,0}=\nu_B\left[3^{\frac{p+1}{2}}\frac{\pi \sqrt \pi}{4}\frac{eRK}{B}f_\kappa(p)\right]^{\frac{2}{p+4}}.
\label{nua0}
\end{eqnarray}

When bursts of a repeating FRB injecting into the nebula, electrons in the nebula would absorb them and a fraction of low-energy electrons would be accelerated to higher energy \citep{yang16}. During this period, the distribution of electrons in the nebula satisfies the continuity equation given by \citep{mcc69}
\begin{eqnarray}
\frac{\partial N_{\rm e}(\gamma,t)}{\partial t}&=&\frac{\partial}{\partial \gamma}\left[A\gamma^2 N_{\rm e}(\gamma,t)\right]+\frac{\partial}{\partial \gamma}\left[C\gamma^2\frac{\partial}{\partial \gamma}\frac{N_{\rm e}(\gamma,t)}{\gamma^2}\right]\nonumber\\
&+&S(\gamma,t).
\label{ke}
\end{eqnarray}
The first term on the right-hand side of Eq.(\ref{ke}) corresponds to the effect of the synchrotron energy loss, with
\begin{eqnarray}
A\gamma^2=\frac{1}{m_ec^2}\int P_{\rm iso}(\nu,\gamma)d\nu,
\end{eqnarray}
where
\begin{eqnarray}
P_{\rm iso}(\nu,\gamma)=\frac{\sqrt{3}e^3B}{m_ec^2}F_{\rm iso}\left(\nu/\nu_{\rm ch}\right)\,\unit {erg\,{cm}^{-2}\,s^{-1}\,Hz^{-1}}\nonumber\\
\end{eqnarray}
is the synchrotron power of a single electron averaging over an isotropic distribution of pitch angle $\alpha$ \citep{wij99}, $\nu_{\rm ch}\equiv(3/2)\gamma^2\nu_B$ is the typical synchrotron characteristic frequency, and the isotropic synchrotron function is
\begin{eqnarray}
F_{\rm iso}(\nu/\nu_{\rm ch})=\int^{\pi/2}_{0}\sin^2\alpha F\left (\frac{\nu}{\nu_{\rm ch}\sin\alpha}\right)d\alpha,
\end{eqnarray}
with
\begin{eqnarray}
F(x)=x\int_x^\infty K_{\frac{5}{3}}(\xi) d\xi.
\end{eqnarray}
The second term on the right-hand side of Eq.(\ref{ke}) corresponds to induced emission and reabsorption, with
\begin{eqnarray}
C=\frac{1}{m_ec^2}\int \frac{I_{\nu,\rm {tot}}}{2m_e\nu^2}P_{\rm iso}(\nu,\gamma)d\nu,\label{cfun}
\end{eqnarray}
where $I_{\nu, \, \rm {tot}}$ is the total intensity, which is the sum of  $I_{\nu,\,\rm{FRB}}$, the effective average intensity of a radio bursts at the nebula, and $I_{\nu,\,\rm {neb}}$, the SSA intensity of the nebula.
For a burst with observed flux $F_\nu$, the burst flux at the nebula is $F_{\nu,\, \rm n}=F_\nu d^2/r^2$, where $r$ is the distance from the FRB to the nebula, and $d$ is the luminosity distance from the FRB to the observer. The luminosity distance of FRB 121102 is estimated to be $d\simeq 3.0\times10^{27}\,\unit{cm}$, when taking the $\Lambda$CDM cosmological parameters as $\Omega_{\rm m}=0.3089\,\pm\,0.0062$, $\Omega_\Lambda=0.6911\,\pm\,0.0062$, and $H_0=67.74\,\pm\,0.46\,\unit{km\,s^{-1}\,Mpc^{-1}}$ \citep{pla16}. Accordingly, at the nebula, the integral effective intensity of a burst of FRB 121102 is~\footnote{Note that in Eq.(\ref{cfun}) the intensity $I_{\nu,{\rm tot}}$ corresponds to a mean intensity that is directly associated with the flux density \citep[e.g.,][]{ryb86}.}
\begin{align}
I_0 &\simeq  \frac{\nu_0 \cdot F_{\nu,\,\rm n}}{\pi}
\simeq \frac{\nu_0 \cdot F_\nu}{\pi}\frac{d^2}{r^2}\simeq 3.0\times 10^7\,\unit{erg\,s^{-1}\,cm^{-2}\,str^{-1}}\notag \\
 &\times(\nu_0/1\unit{GHz})(F_\nu/1\,\unit{Jy})(r/0.01\,\unit{pc})^{-2},
\end{align}
where $\nu_0$ is the characteristic frequency of a burst.
The integral synchrotron intensity of the nebula is
\begin{eqnarray}
\nu_a I_{\nu_a,\,\rm{neb}}&\simeq 2.6\times10^4\,\unit{erg\,s^{-1}\,cm^{-2}\,str^{-1}}\notag \\
&\times(L/10^{39}\unit{erg\,s^{-1}})(r/0.01\,\unit{pc})^{-2},
\end{eqnarray}
where $L$ is the luminosity of the nebula.
Since $I_0\gg\nu_aI_{\nu_a,\,\rm{n}}$, then $I_{\nu,\,\rm{tot}}$ can be approximated to be $I_{\nu,\,\rm{FRB}}$.
Since the spectra of the bursts of FRB 121102 are narrow~ \footnote{According to \citet{law17}, the typical spectra of some bursts of FRB 121102 are with center frequency of $\sim 3\,\unit{GHz}$ and FWHM of $\sim 500\,\unit{MHz}$, which are narrow spectra.}, we assume that the spectrum of a single burst is a $\delta$-function, i.e., $I_{\nu,\,\rm{FRB}}=I_0\delta(\nu-\nu_0)$, then the coefficient $C$ can be approximately written as
\begin{eqnarray}
C\simeq \frac{I_0}{2m_e^2c^2\nu_0^2}P_{\rm iso}(\nu_0,\gamma).
\end{eqnarray}
The last term $S(\gamma,t)$ on the right-hand side of Eq.(\ref{ke}) represents electron injection in the emission region.
We assume that the electron injection is zero, e.g., $S(\gamma,t)=0$.
During the burst injection, the intensity of the nebula is given by
\begin{eqnarray}
I_\nu^\prime=\frac{j'_\nu R}{\tau_\nu^\prime}(1-e^{-\tau_\nu^\prime}),
\label{Itot}
\end{eqnarray}
with
\begin{eqnarray}
j_\nu^\prime&=&\frac{1}{4\pi}\int_{\gamma_{\min}}^{\gamma_{\max}} N_{\rm e}(\gamma,t)P_{\rm iso}(\nu,\gamma)d\gamma,
\end{eqnarray}
and
\begin{eqnarray}
\tau_\nu^\prime&=&\frac{R}{8\pi m_e\nu^2}\int_{\gamma_{\min}}^{\gamma_{\max}}\frac{N_{\rm e}(\gamma,t)}{\gamma^2}\frac{d}{d\gamma}[\gamma^2 P_{\rm iso}(\nu,\gamma)]d\gamma.
\end{eqnarray}
By numerically solving Eq. (\ref{ke}), we can figure out the electron distribution and the spectrum of the nebula at any time. Finally the flux of the nebula is given by
\begin{eqnarray}
F_\nu^\prime \simeq \pi I_\nu^\prime\frac{r^2}{d^2}.
\end{eqnarray}

In the following discussion, we will present the nebula spectrum via the numerical method after considering that the nebula is heated by multiple bursts of a repeating FRB source, and constrain the model parameters with the observations of FRB 121102.

\section{Numerical calculation results}\label{sec3}

In order to calculate the spectrum of FRB-heated nebula and perform numerical solutions, we make an assumption that the flux, the frequency and  the duration  of a burst are independent of each other. \cite{gaj18} and \cite{zha18} identified 93 bursts of FRB 121102 during 5 hours by GBT at 4-8 GHz. This observation composes the largest sample of FRB 121102 for a single continuous observation up to now.  Based on this observation, \cite{zha19} analyzed power-law distributions for energies, fluxes, durations and waiting times of FRB 121102. They found that the distribution of fluxes meets $dN/dF_{\nu}\varpropto F^{\alpha}_{\nu}=F^{-1.94}_{\nu}$ with $F_{\nu}$ from $F_{\nu,\min}=0.001\,\unit{Jy}$ to $F_{\nu,\max}=1\,\unit{Jy}$, the distribution of durations is about $dN/dt\varpropto t^{\gamma}=t^{-1.57}$ with $t$ ranging from $t_{\min}=0.7\,\unit{ms}$ to $t_{\max}=4\,\unit{ms}$, and the average of waiting times $T$ is around $100\,\unit{s}$. Here the waiting time is defined as the difference of occurring time for two bursts. We assume that the intrinsic frequency distribution of FRB 121102 also satisfies the form of a power law, which reads $dN/d\nu_0\varpropto \nu^{\beta}_0$ with the frequency of FRB 121102 from $\nu_{0,\min}=0.1\,\unit{GHz}$ to $\nu_{0,\max}=8\,\unit{GHz}$ (due to the nebula absorption, some low-frequency busts would be unobservable).
We further suppose that the distributions of flux and duration can be extended to low frequency. Based on the above assumptions, the total number of samples is
\begin{eqnarray}
N_{\rm total}=\int_{F_{\nu,\min}}^{F_{\nu,\max}}\int_{\nu_{0,\min}}^{\nu_{0,\max}}\int_{t_{\min}}^{t_{\max}}\frac{dN}{dF_{\nu}d\nu_0 dt}dF_{\nu}d\nu_0 dt,\nonumber\\
\end{eqnarray}
and the total injection energy is
\begin{align}
E_{\rm total}&\simeq \frac{4\pi d^2}{1+z}\int_{F_{\nu,\min}}^{F_{\nu,\max}}\int_{\nu_{0,\min}}^{\nu_{0,\max}}\int_{t_{\min}}^{t_{\max}}F_{\nu}\delta\nu_0  \delta t \frac{dN}{dF_{\nu}d\nu_0 dt}dF_{\nu}d\nu_0 dt\notag\\
&\simeq \frac{4\pi d^2}{1+z}\int_{F_{\nu,\min}}^{F_{\nu,\max}}\int_{\nu_{0,\min}}^{\nu_{0,\max}}\int_{t_{\min}}^{t_{\max}}F_{\nu} \nu_0  t \frac{dN}{dF_{\nu}d\nu_0 dt}dF_{\nu}d\nu_0 dt,
\label{Etotal}
\end{align}
where $z =0.193$ is the redshift of FRB 121102 \citep{ten17}, and $\delta \nu_0$ and $\delta t$ are widthes of the spectrum and the pulse, respectively. At last, we assume that $\gamma_{\min}=10$ and $\gamma_{\max}=10^5$ for the electron distribution satisfying Eq.(\ref{electrondistribution}).

\begin{figure}[]
\centering
\includegraphics[angle=0,scale=0.5]{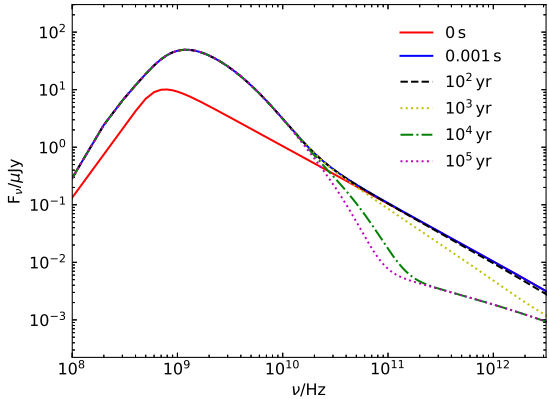}
\caption{The spectra of the nebula before and after a burst injection. The solid red line denotes the initial spectrum of the nebula without any burst injection. The solid blue line denotes the spectrum of nebula with one burst injection. The model parameters are taken as $p=3$, $\nu_0=0.3\, \unit{GHz}$, $E=10^{49}\, \unit{erg}$, $t=0.001\,\unit{s}$, $\nu_{\rm a,0}=0.6\,\unit{GHz}$, $r=0.03\,\unit{pc}$, and $R=10^{13}\,\unit{cm}$. The other lines indicate evolutions on the spectra of the nebula after the burst injection, from one hundred years to one hundred thousands years, respectively.
} \label{fig1}
\end{figure}

First, we compare the nebula spectra before and after a radio burst injection, as shown in Figure~\ref{fig1}.
We find that there is no significant evolution on the spectrum of the nebula after the burst injection.
The reason is as follows: the synchrotron heating would cause a harder electron spectrum with a peak Lorentz factor of
\begin{align}
\gamma_{\rm peak}\sim(\nu_0/\nu_B)^{1/2}\sim 710(B/1\,\unit{mG})^{-1/2}(\nu_0/1.4\,\unit{GHz})^{1/2}
\label{gammapeak}
\end{align}
in the nebula \citep[see Figure~1 and Figure~2 in][]{yang16}, where $\nu_0$ is the burst frequency, and $\nu_B$ is the Larmor frequency with magnetic field strength $B \simeq 1\,\unit{mG}$ (inferred by the RM of FRB 121102, \citet{mic18}).
For the nebula by synchrotron radiation, the cooling time scale $\delta t_{\rm cooling}$ satisfies
\begin{align}
\delta t_{\rm cooling}&\simeq \left(\dfrac{2e^4B^2}{3m_e^3c^5}\, \gamma_{\rm peak}\right)^{-1}\sim \dfrac{3m_e^3c^5}{2e^4B^2}\, \left(\frac{\nu_0}{\nu_B}\right)^{-1/2}\notag \\
&\sim2.3\times10^4\,\unit{yr}~(B/1\,\unit{mG})^{-3/2}(\nu_0/1.4\,\unit{GHz})^{-1/2}.
\end{align}
According to the observation of FRB 121102 \citep{gaj18,zha18}, the average waiting time between bursts is $T\sim100\,\unit s\ll\delta t_{\rm relax}$, hence the impact of synchrotron cooling on the nebula spectrum is very slight. Therefore, it can be considered that the nebula spectrum does not change within a few hundred seconds of two burst injections.

As shown in Figure~\ref{fig1}, we compare the nebula spectra which are before a burst injection, with one burst injection and after this injection for a long-term evolution, $10^2$, $10^3$, $10^4$, and $10^5$ years, respectively.
The difference of the nebula spectra between one burst injection without cooling  and with $10^5$-year cooling is less than 21$\%$ in the GHz band.
Notice that, in Figure~\ref{fig1}, in order to make the synchrotron heating effect appear significant, we take the energy of a injecting burst as~\footnote{For FRB 121102, the typical energy of a single burst is about $E\sim10^{38}~\unit{erg}$, thus the synchrotron heating effect is not significant for such a single burst. However, as lots of bursts injecting into the nebula, the nebula spectrum would change significantly, which is discussed later.} $E=10^{49}\,\unit{erg}$. The other model parameters are taken as the index of the initial electron spectrum in the nebula, $p=3$, the initial SSA frequency of the nebula, $\nu_{\rm a,\,0}=0.6\,\unit{GHz}$, the frequency of the burst, $\nu_0=0.3\,\unit{GHz}$, the thickness of the nebula, $R=10^{13}\,\unit{cm}$, the distance from the burst to the nebula, $r=0.03\,\unit{pc}$, and the duration of the burst, $t=0.001\,\unit{s}$. The energy loss of the nebula through the synchrotron radiation during a long time is not significant in the GHz band that is we concerned. Thus, in the following discussion, we only consider the spectrum evolution of the nebula in the process of burst injections, and ignore the evolution during every waiting time.

\begin{figure}[]
\centering
\includegraphics[angle=0,scale=0.5]{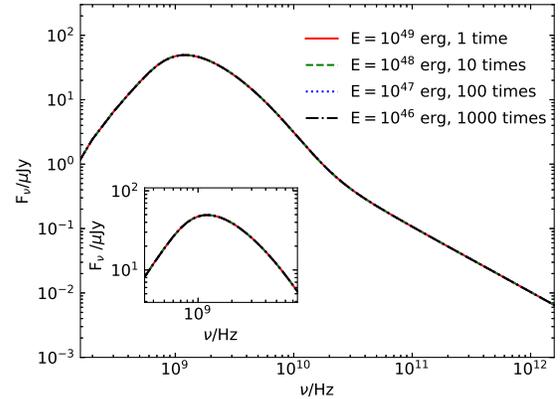}
\caption{The spectra of the nebula with the same total energy of bursts injecting at the frequency $\nu_0=0.3\,\unit{GHz}$. $E$ is the energy of a single burst and the following number represents the total number of injections. Different lines denote situations with different energies of each burst and times of injections, but the same total energy. The model parameters are taken as $p=3$, $t=0.001\,\unit{s}$, $\nu_{\rm a,0}=0.6\,\unit{GHz}$, $r=0.03\,\unit{pc}$, and $R=10^{13}\,\unit{cm}$. The insert is the amplification at GHz. Above parameters,  like $E$ and injection times, are chosen only to illustrate the results so may not match the actual.} \label{fig2}
\end{figure}

Next, we discuss the synchrotron heating effect of multiple injections at a certain frequency. As shown in Figure~\ref{fig2}, for a given burst frequency,
if the total injection energies are identical, multiple low-energy injections can be combined into one high-energy injection.
To illustrate this, we consider several cases. For example~\footnote{Similar to the above discussion, here a large energy for a single burst is taken in order to make synchrotron heating effect significant.}, one-thousand burst injections, each with the energy of $10^{46}\,\unit{erg}$ and the frequency of $0.3\,\unit{GHz}$, can have the same effect on the nebula compared with only one burst injection with the energy of $10^{49}\,\unit{erg}$ at the same frequency and their difference is less than $0.01\%$. In the model, the parameters are chosen as the duration of each burst, $t=0.001\,\unit{s}$, the index of the initial electron spectrum in the nebula, $p=3$, the initial SSA frequency of the nebula, $\nu_{\rm a,0}=0.6\,\unit{GHz}$, the distance from bursts to the nebula, $r=0.03\,\unit{pc}$, and the thickness of the nebula, $R=10^{13}\,\unit{cm}$. It can be interpreted as that the effects on the spectra of nebula mainly depend on the total energy injection at that frequency, not the number of bursts.
Therefore, we can combine the bursts with the same frequency into one higher-energy injection.
In the following simulations, we combine $10^7$ bursts into one high-energy burst, i.e., the energy of every burst is $10^7$ times of the real one, and a large population of bursts are also needed.

\begin{figure}[]
\centering
\includegraphics[angle=0,scale=0.5]{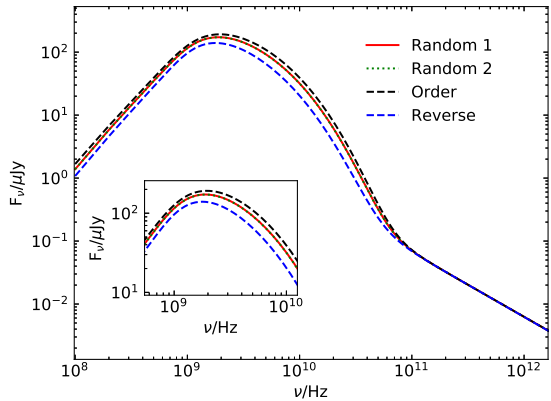}
\caption{The spectra of the nebula obtained from different injection sequences of the same set of bursts. The ``Random 1'' and ``Random 2'' lines denote the injections with two different disordered sequences, respectively. The dashed black line, ``Order'', denotes the frequencies of the bursts from low to high. On the contrary, the blue dashed line,  ``Reverse'', corresponds to the frequencies of the burst from high to low. The insert is the amplification at GHz. The parameters used in this group of samples meet $p=3$, $E_{\rm total}=10^{49}\, \unit{erg}$, $\nu_{\rm a,0}=0.52\,\unit{GHz}$, $\beta=-3$, $r=0.03\,\unit{pc}$, and $R=10^{13}\,\unit{cm}$.} \label{fig3}
\end{figure}

We analyze the synchrotron heating effects of the burst injections with different frequency sequences.
We consider different injection sequences of these samples in frequency, which are ``Random'', ``Order''(from low frequency to high frequency), and ``Reverse''(from high frequency to low frequency). In terms of different disordered sequences, there is almost no distinction on the spectra of the nebula with the difference less than $0.6\%$, as shown in Figure~\ref{fig3}.
Either ``Order'' or ``Reverse'' is a special case which has the difference less than $40.5\%$ from the situation of ``Random''.
This group of samples meets the total energy of the bursts, $E_{\rm total}=10^{49}\, \unit{erg}$, and the index of frequency distribution, $\beta=-3$. Other model parameters are chosen as $p=3$, $\nu_{\rm a,0}=0.52\,\unit{GHz}$, $r=0.03\,\unit{pc}$, and $R=10^{13}\,\unit{cm}$. In the case where a series of bursts with different fluxes, durations and frequencies inject into the nebula, the order of the burst injections has little effects on the final spectra of the nebula. Therefore, we ignore the impact of the sequence of the injection in the following simulations and we will simulate the spectra of the nebula with samples out of order in frequency.

\begin{figure}[]
\centering
\includegraphics[angle=0,scale=0.5]{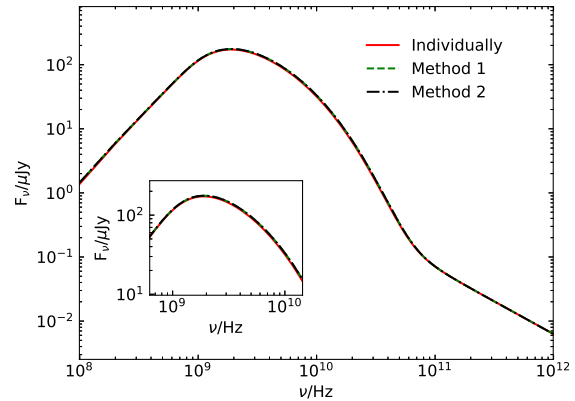}
\caption{The spectra of the nebula obtained by different methods. The red line denotes a group of bursts injecting into the nebula individually. Method 1 is to combine the energies of these bursts in the unit frequency interval and the combined bursts are disordered. Method 2 is to use Eq.(\ref{Etotal}) to calculate the energy of these bursts in the unit frequency interval and the injection is disordered in frequency. The insert diagram enlarges the fluxes of three methods in the range of $1\,\unit{GHz}$ to $20\,\unit{GHz}$. These three methods adopt parameters, which are $p=3$, $E_{\rm total}=10^{49}\, \unit{erg}$, $\nu_{\rm a,0}=0.52\,\unit{GHz}$, $\beta=-3$, $r=0.03\,\unit{pc}$, and $R=10^{13}\,\unit{cm}$.} \label{fig4}
\end{figure}

To simplify the simulation process, we evaluate two methods to obtain the nebula spectra. The first method is that after generating a population of samples with a total energy and an index of frequency distribution of these bursts, we combine the bursts in the unit frequency interval, which can be regarded as bursts with the same frequency joined together (see Figure~\ref{fig2}). The second method is that without generating samples, we directly calculate the energy injecting into the nebula in the unit frequency interval by Eq.(\ref{Etotal}). As shown in Figure~\ref{fig4}, the spectra of the nebula obtained by these two methods are similar to the method with bursts injecting individually with the difference less than $5.6\%$.
The first method greatly reduces the number of injections that needs to be simulated.
The second method does not need to produce a large amount of bursts, but the injection energy of per unit frequency is directly obtained.
So in the next simulations, we mainly use the second method.
These methods adopt the parameters, $p=3$, $E_{\rm total}=10^{49}\, \unit{erg}$, $\nu_{\rm a,0}=0.52\,\unit{GHz}$, $\beta=-3$, $r=0.03\,\unit{pc}$, and $R=10^{13}\,\unit{cm}$.

\begin{figure}[]
\centering
\includegraphics[angle=0,scale=0.285]{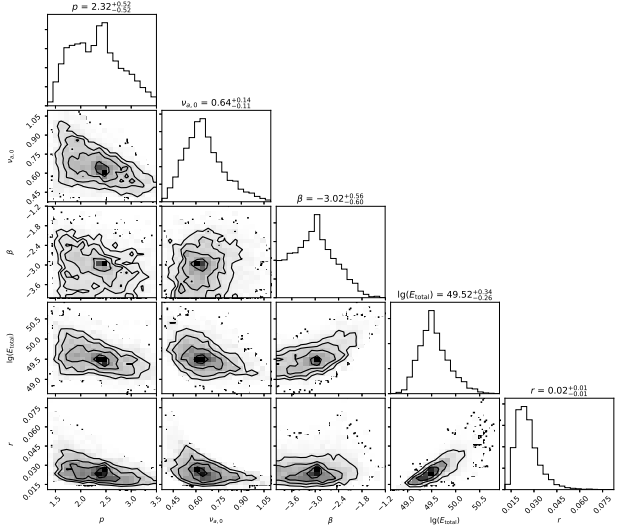}
\caption{The contour lines of constraining model fitting parameters at 0.5$\sigma$, 1$\sigma$, 1.5$\sigma$, and 2$\sigma$ significance levels. The best fitting values are shown in the figure. There are five parameters restricted in our model, which are the index of the initial electron spectrum in the nebula, $p$, the initial SSA frequency of the nebula, $\nu_{\rm a,\,0}$, the index of frequency distribution of bursts, $\beta$, the total energy of the bursts,  $E_{\rm total}$, and the distance from the bursts to the nebula, $r$, respectively.} \label{fig5}
\end{figure}

\begin{figure}[]
\centering
\includegraphics[angle=0,scale=0.5]{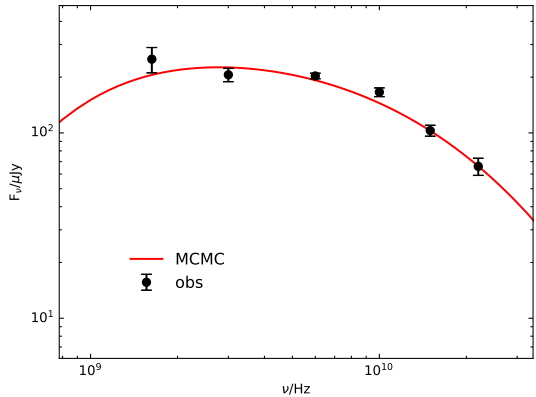}
\caption{The spectra of the nebula with the best fitting parameters according to the MCMC method and from the observation \citep{cha17}. The red solid line denotes the spectrum of the nebula corresponding to the best fitting parameters, and the black points represent the observation values. Error bars represent 1$\sigma$ uncertainties.} \label{fig6}
\end{figure}

We apply the Markov Chain Monte Carlo (MCMC) method to constrain our model parameters with the code emcee \citep{for13}. The log likelihood for these parameters can be determined by a $\chi^2$ statistic, i.e.,
\begin{eqnarray}
\chi^2 (p, \nu_{\rm a,0},\beta , E_{\rm total}, r)=\sum\limits_{i} \frac{(F_{{\rm{obs}},  i}-F_{\nu,i})^2}{\sigma^2_{{\rm{obs}},i}},
\end{eqnarray}
where $i$ represents different data at different frequencies, $F_{{\rm{obs}},  i}$ is the observed flux \citep{cha17}, $F_{\nu,i}$ is the flux obtained according to given model parameters at the frequency of the observation by interpolation, and $\sigma^2_{{\rm{obs}},i}$ is the corresponding error of the observation. By calculating and minimizing the $\chi^2$ for a wide range of the parameters of the model and converting each $\chi^2$ into the log-probability function, we obtain the projections of the posterior probability distribution of the fitting parameters in the contours, as shown in Figure~\ref{fig5}. Moreover, the best fit to the observed spectrum of the nebula is shown in Figure~\ref{fig6}. The parameters restricted in our model are the index of the initial electron spectrum in the nebula, $p$, the initial SSA frequency of the nebula, $\nu_{\rm a,\,0}$, the index of frequency distribution of bursts, $\beta$, the total energy of the bursts,  $E_{\rm total}$, and the distance from the bursts to the nebula, $r$. Since $\nu_{\rm a,\,0}$, $K$ and $R$ are not independent, we can derive the product of $K$ and $R$ by Eq.(\ref{nua0}).

The analysis results show that the initial electron spectrum index is $p=2.32^{+0.52}_{-0.52}$, the initial SSA frequency of nebula is $\nu_{\rm a,0}=0.64^{+0.14}_{-0.11}\,\unit{GHz}$, the index of the frequency distribution of bursts satisfies $\beta = -3.02^{+0.56}_{-0.60}$, the total energy of injecting bursts should meet $\lg (E_{\rm total}/\,\unit{erg})=49.52^{+0.34}_{-0.26}$, and the radius of the nebula is $r=0.02^{+0.01}_{-0.01}\,\unit{pc}$.
Then the total number of the bursts injecting into the nebula is around $N_{\rm total}\simeq 2.26\times10^{12}$.
At present, the observations and statistics of FRB 121102 are concentrated on $\gtrsim1\,\unit{GHz}$ \citep[e.g.,][]{spi16,sch16,law17,mic18,gaj18,spi18,zha18,hes19,gou19} with the average of waiting times is about $100\,\unit{s}$ \citep{zha19}. By the best fit, We also obtain that the number of bursts above $1\,\unit{GHz}$ is $N_{\rm total}(>1\,\unit{GHz})\simeq2.11\times10^{10}$. Without considering the connection between burst fluence and waiting time \citep{gou19}, we can estimate the age of the nebula is more than $N_{\rm total}(>1\,\unit{GHz})\times 100\,\unit{s} \sim 6.7\times 10^4\,\unit{yr}$.
If the thickness of the nebula is $R=10^{13}\,\unit{cm}$,  according to Eq.(\ref{nua0}) we can get $K \simeq 1.8\times10^9\,\unit{cm^{-3}}$. So the total number of electrons in the nebula is about $3.2\times10^{55}$.

\begin{figure}[]
\centering
\includegraphics[angle=0,scale=0.5]{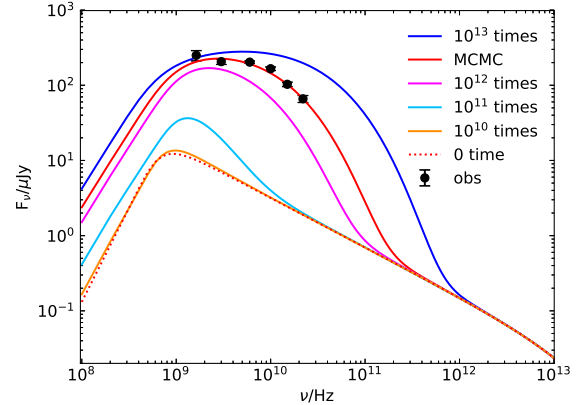}
\caption{The evolution of the spectra of the nebula with injection times. Different lines correspond to different injection times and black points represent the observation values of the persistent emission of FRB 121102 \citep{cha17}. In addition to different injection times (i.e., different injection energy), the other model parameters are the best fitting parameters obtained by the MCMC method. } \label{fig7}
\end{figure}

In order to study the evolution of the spectra of the nebula with injection times, we simulate the process of bursts injecting into the nebula using the best fit parameters of the MCMC method but with different injection times, as shown in Figure~\ref{fig7}.
We find that the spectra of the nebula with burst injections of $10^{10}$ times is basically the same as that without burst injections, in which case the synchrotron heating effect is not significant.
As the burst injections increasing, the peak frequency and the peak flux of the nebula would increase.
Once the injection times exceed $2\times10^{12}$ times, e.g., $6.7\times10^4\,\unit{yr}$ since the first burst injection, the predicted brightness of the nebula would be higher than the current observation, which might imply that the persistent emission of FRB 121102 would be brighter in the future.

\begin{figure}[]
\centering
\includegraphics[angle=0,scale=0.5]{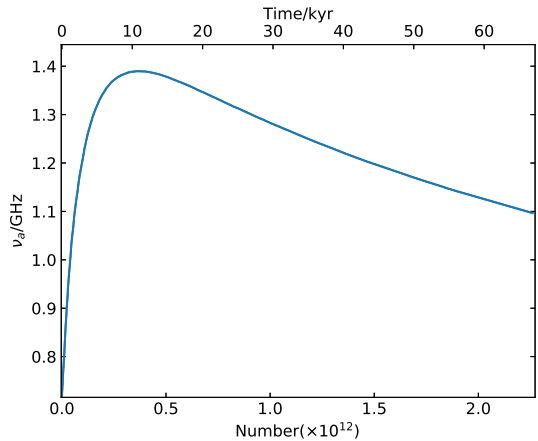}
\caption{The evolution of the SSA frequency of the nebula with the best fitting parameters obtained by the MCMC method. The number indicates how many burst injections, and the time indicates how long does it take. According to the observations and statistics of FRB 121102 concentrated on over $1\,\unit{GHz}$, the average of waiting times between bursts is about $100\,\unit{s}$. The maximum value of the time is obtained by $100\,\unit{s}\times N_{\rm total}(>1\,\unit{GHz})$. Besides, assuming that the waiting time between every two bursts is the same, the time interval is derived by averaging the maximum value of the time with the total number of bursts. This is a correspondence between the number of burst injections and the time of the injection process with a mean waiting time considered. } \label{fig8}
\end{figure}

Finally, we consider the evolution of the SSA frequency $\nu_{\rm a}$ of the nebula during burst injecting separately. We consider the number of burst injections with the time of the injection process according to the mean waiting time, and assume that the waiting time between every two bursts is the same as the one that is derived by dividing the estimated age of the nebula by all the number of bursts. Because the electron distribution of the nebula will change gradually with the energy of each burst injection, in the process of burst injections, the SSA frequency of the nebula is not fixed at $\nu_{\rm a,0}$. The final SSA frequency is larger than the initial one.
As shown in Figure~\ref{fig8}, we can see that $\nu_{\rm a}$ increases first and then decreases with burst injections, and is currently maintained near $1.1\,\unit{GHz}$. This is consistent with observations that the repeating bursts of FRB121102 are mainly observed in the GHz band. The bursts at $\nu\,<\,\nu_a$ are absorbed and used to accelerate electrons in the nebula.

\begin{figure*}[]
\centering
\includegraphics[angle=0,scale=0.5]{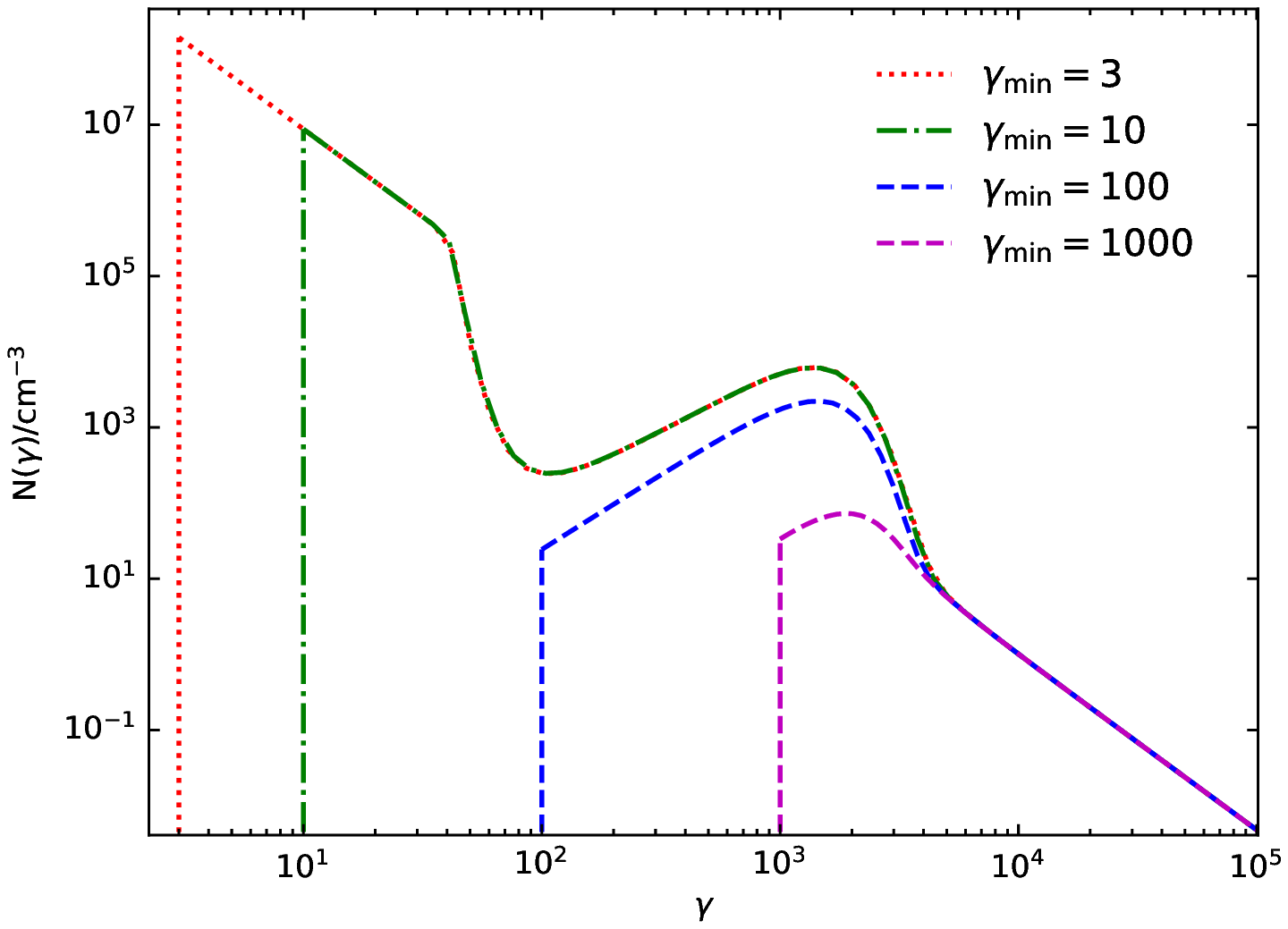}
\includegraphics[angle=0,scale=0.5]{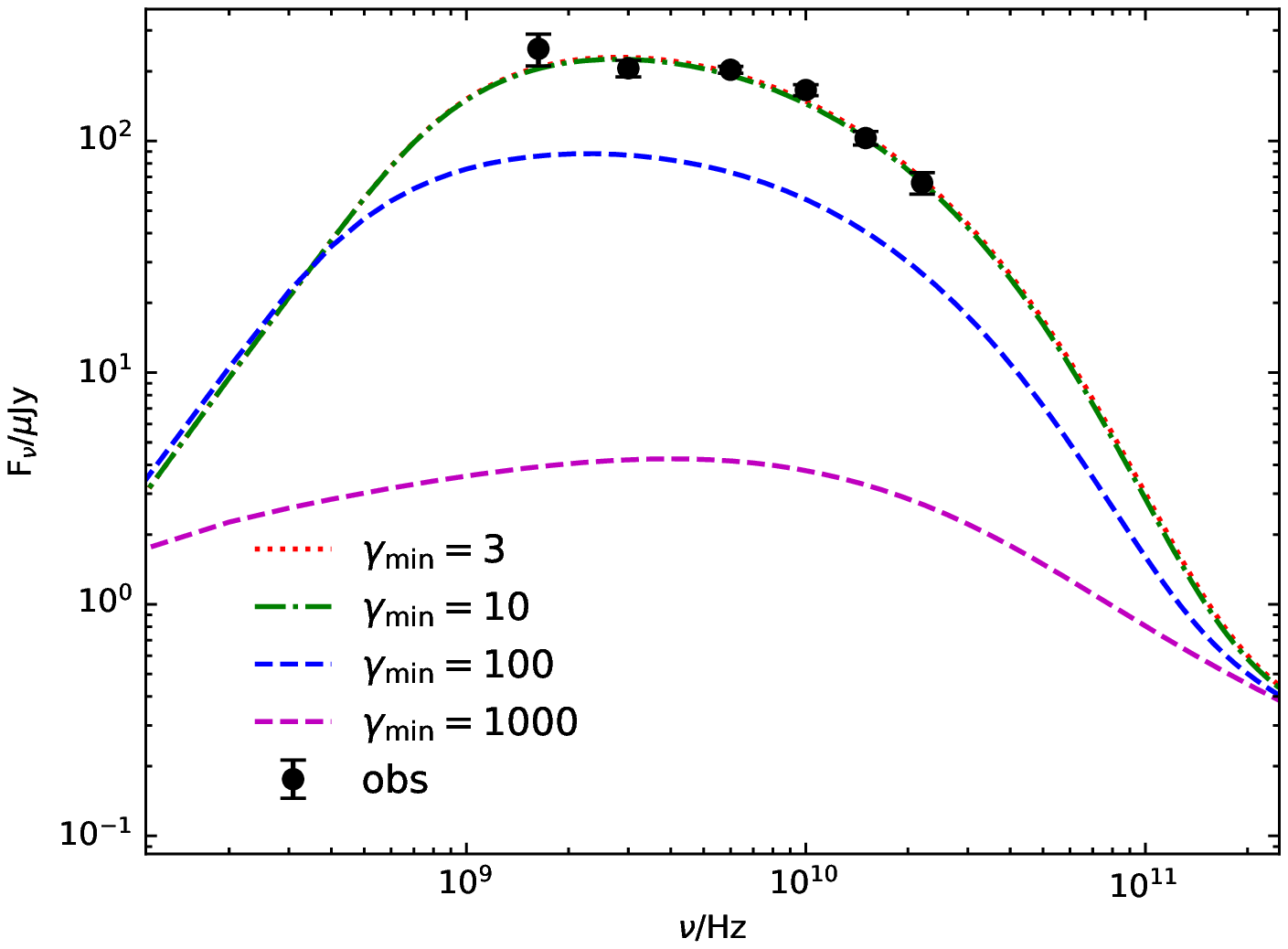}
\caption{Left panel: the electron spectra. Right panel: the nebula emission spectra with the observation data \citep{cha17}. The minimum electron Lorentz factor is taken as $\gamma_{\rm min}=3,10,100,1000$.
The red dotted lines denote $\gamma_{\rm min}=3$, the green dash-dotted lines denote $\gamma_{\rm min}=10$, the blue dashed lines denote $\gamma_{\rm min}=100$, and the violet dashed lines denote $\gamma_{\rm min}=1000$.
Besides $\gamma_{\rm min}$, the other model parameters are the best fitting parameters obtained by the MCMC method. In the right panel, the red dotted line and green dash-dotted line are overlapping.} \label{fig9}
\end{figure*}

In the above simulations, the minimum electron Lorentz factor is taken as $\gamma_{\rm min}=10$. For a wide range of $\gamma_{\rm min}$, we find that only if $\gamma_{\rm min} \ll \gamma_{\rm peak}$, where $\gamma_{\rm peak}$ is the electron peak Lorentz factor given by Eq.(\ref{gammapeak}), the synchrotron heating effect is not significantly depend on $\gamma_{\rm min}$, because only electrons at $\gamma_{\rm peak}$ are significantly accelerated by FRBs. For $\gamma_{\rm min} \gg \gamma_{\rm peak}$, the synchrotron heating effect is unimportant.
In this model with the above parameters, the peak Lorentz factor is $\gamma_{\rm peak}\simeq700$.
In Figure~\ref{fig9}, we plot the electron spectra (left panel) and nebula spectra (right panel) with different minimum Lorentz factors, $\gamma_{\rm min}=3,10,100,1000$. We find that the nebula spectra show a difference less than 3.1\% in terms of $\gamma_{\rm min}$ taken as 3 and 10. The difference between situations of $\gamma_{\rm min}=10$ and $\gamma_{\rm min}=100$ is less than 62.8\%. When $\gamma_{\rm min}=1000$, the spectrum of the nebula is very different from that of $\gamma_{\rm min} \ll \gamma_{\rm peak}$.
On the other hand, in the scenario we discuss here, the value of $\gamma_{\rm min}$ is determined by the initial electron injection and synchrotron cooling together, i.e., $\gamma_{\rm min}=\min(\gamma_{\rm min,0},\gamma_c)$, where $\gamma_{\rm min,0}$ is the minimum Lorentz factor of initial injection electrons and $\gamma_c$ is the Lorentz factor due to synchrotron cooling. At the observation time $t\simeq6.7\times 10^4\,\unit{yr}$, the synchrotron cooling leads to $\gamma_{\rm min}\leqslant\gamma_c\simeq6\pi m_e c/(\sigma_T B^2 t)\simeq 370$ for $B\sim1\,\unit{mG}$. Therefore, a relatively low value of $\gamma_{\rm min}$ we adopt above is reasonable.
At last, based on the best fit of the MCMC, the electron number in the nebula is about $3.2\times10^{55}(\gamma_{\rm min}/10)^{-1.32}$, so the corresponding mass of the nebula is about $0.03M_{\odot}(\gamma_{\rm min}/10)^{-1.32}$ for baryon-dominated ejecta or $10^{-5}M_{\odot}(\gamma_{\rm min}/10)^{-1.32}$ for pair-dominated ejecta.

\section{Conclusions and Discussion}\label{sec4}

In this paper, we have discussed the physical process of synchrotron heating by multiple bursts of a repeating FRB in a nebula, and used this model to explain the persistent radio emission of FRB 121102 \citep{cha17}.
Due to the synchrotron heating effect, the electron distribution in the nebula will change gradually, As a result, the spectrum of the nebula will show a significant evolution at the SSA frequency.
Different from a non-repeating FRB injecting into a nebula \citep{yang16}, the multiple bursts from a repeating FRB source could provide a tremendous amount of energy to the nebula, and finally transfer to the observed persistent emission of the repeating FRB.
For FRB 121102, the lowest frequency of the detected bursts is about 600 MHz \citep{jos19}.
Our results show that the bursts with frequency lower than the SSA frequency $\nu_{\rm a}\sim 0.6~{\rm GHz}$ of the nebula would be absorbed. Since the spectrum of a single burst is narrow, most low-frequency bursts with frequency lower than $\nu_{\rm a}$ is completely absorbed, and only the bursts with frequency closed to $\nu_{\rm a}$ are partly absorbed, leading to harder spectra. Meanwhile, the frequency distribution of bursts would also become harder after absorption.

In the above analysis, we assumed that the flux, the frequency and the duration of a burst are independent of each other, and they satisfy power-law distributions \citep{zha19}.
Consider some low-frequency bursts would be absorbed, the intrinsic burst frequencies are assumed to be ranging from $0.1\,\unit{GHz}$ to $8\,\unit{GHz}$.
After a burst injecting into the nebula, the electron distribution in the nebula would change due to synchrotron heating effect and synchrotron cooling together. Since the cooling time scale of the synchrotron radiation in the nebula is about thousands of years, there is no evolution on the spectrum of the nebula during the waiting time of two bursts. Thus it is reasonable that we could ignore the evolution on the spectrum during each time interval of burst injections.

Through numerical calculation, we found that at a certain frequency, multiple low-energy injections could be approximately equivalent to one high-energy injection, if the total injecting energies are identical. On the other hand, for the bursts with different frequencies, the order of the burst injections has little effects on the final spectra of the nebula.
At last, the method directly using Eq.(\ref{Etotal}) to calculate the energy injecting into the nebula in the unit frequency interval, can achieve the similar spectra of the nebula as the method that bursts inject separately.
With the MCMC method to restrict our model parameters, the analysis results show that the total energy of injecting bursts should meet $\lg (E_{\rm total}/\,\unit{erg})=49.52^{+0.34}_{-0.26}$, the age of the nebula is more than $6.7\times 10^4\,\unit{yr}$, and the electron number in the nebula is about $3.2\times10^{55}(\gamma_{\rm min}/10)^{-1.32}$.

In the above discussion, we assume that the electron injection in the nebula is zero for a closed system. If the nebula is shocked by the ambient medium, the electron injection from a shock wave would be considered. However, since the particle injection term (the third term) is much less than the heating term (the second term) in the kinetic equation given by Eq.(\ref{ke}) for the parameters we discussed here, the particle injection could be neglected. Therefore, the election injection plays an insignificant role in this model.

In our model, as the burst injection time increases, the peak frequency and peak flux of the nebula can increase continuously. Meanwhile, according to this model, for non-repeating FRBs, since the total injection energy is much smaller than that of the repeating one, their persistent emission would be not observable, which is consistent with the current observation \citep{mah18}.
For the persistent emission of FRB 121102, the best fit gives a total number of the bursts injecting into the nebula of $N_{\rm total}\simeq 2.26\times10^{12}$ (much larger than the burst number above $1\,\unit{GHz}$, e.g., $N_{\rm total}(>1\,\unit{GHz})\simeq2.11\times10^{10}$), which means that there should be one burst per second in average.
We predicted that as more bursts inject, the persistent emission of FRB 121102 would be brighter than the current observation in the future.
On the other hand, the waiting time could be below 1\,s according to a study of 41 low $S/N$ FRB 121102 bursts \citep{gou19}. It is suggested that some bursts may be missed due to the limited observing bandwidth or the detection threshold for the faint bursts.
We herein present another potential that the fluxes of some low-frequency bursts would be absorbed more or less if their frequencies are below the SSA frequency of the nebula. In the simulation, the SSA frequency of the nebula is currently maintained near $1.1\,\unit{GHz}$, which could explain why most bursts of FRB 121102 were observed at $\nu\gtrsim1\,\unit{GHz}$.

\acknowledgments
We thank an anonymous referee for providing helpful comments and suggestions. This work was supported by the National Key Research and Development Program of China (grant No. 2017YFA0402600) and the National Natural Science Foundation of China (grant Nos. 11573014 and 11833003).

\end{document}